\documentstyle[12pt]{article}

\begin{document}

\newcommand{\beq}{\begin{eqnarray}}
\newcommand{\eeq}{\end{eqnarray}}
\newcommand{\nn}{\noindent}
\newcommand{\non}{\nonumber}
\newcommand{\ee}{e^+ e^-}
\newcommand{\ra}{\rightarrow}
\newcommand{\lra}{\longrightarrow}
\newcommand{\s}{\\ \vspace*{-2mm} }
\newcommand{\sx}{\\ \vspace*{-3mm} }
\newcommand{\Lw}{{\rm log} \frac{1-\mu^2 +w}{w} }
\newcommand{\Lz}{{\rm log} \frac{1-\mu^2 +z}{z} }

\renewcommand{\thefootnote}{\fnsymbol{footnote} }

\nn \hspace*{12cm} UdeM-LPN-TH-93-158 \\
\hspace*{12cm} July 1993 \\

\vspace*{1.8cm}

\centerline{\large{\bf New Fermions at e$^+$e$^-$ Colliders:}}

\vspace*{5mm}

\centerline{\large{\bf II.~Signals and Backgrounds.}}

\vspace*{1.5cm}

\centerline{\sc G.~Azuelos$^{1,2}$ and A.~Djouadi$^1$\footnote{NSERC Fellow.}.}

\vspace*{1cm}

\centerline{$^1$ Laboratoire de Physique Nucl\'eaire, Universit\'e de
Montr\'eal,  Case 6128 Suc.~A,}
\centerline{H3C 3J7 Montr\'eal PQ, Canada.}
\vspace*{2mm}
\centerline{$^2$ TRIUMF, 4004 Wesbrook Mall, V6T 2A3 Vancouver BC, Canada.}

\vspace*{2cm}

\begin{center}
\parbox{14cm}
{\begin{center} ABSTRACT \end{center}
\vspace*{0.2cm}

\nn We discuss the production, at high--energy e$^+$e$^-$ linear colliders, of
new heavy fermions predicted by extensions of the Standard Model. We analyze
in great details the various signals and the corresponding backgrounds for both
pair production and single production in association with ordinary fermions.
Concentrating on new leptons, we use a model detector for e$^+$e$^-$ collisions
at a center of mass energy of 500 GeV, to illustrate the discovery potential of
the Next Linear Colliders.}

\end{center}

\renewcommand{\thefootnote}{\arabic{footnote} }
\setcounter{footnote}{0}

\newpage

\subsection*{1.~Introduction}

\nn Many theories beyond the Standard Model [SM] of the electroweak and strong
interactions predict the existence of new fermions. In most of the cases, these
new fermions have non--canonical ${\rm SU(2)_L \times U(1)_Y}$ quantum numbers,
e.g.~the left--handed components are in weak isosinglets and the right--handed
components in weak isodoublets. Examples of new fermions are the following
(for a review, see for instance Ref.~\cite{R1}): \s

\nn $i)$  Sequential fermions: this is the simplest extension of the SM, one
simply has to add to the known fermionic spectrum with its three--fold replica
a fourth family with the same quantum numbers. The existence of this fourth
generation is still allowed by available experimental data, if the associated
neutrino is heavy enough \cite{R2}. \sx

\nn $ii)$ Vector fermions: these occur for instance in the exceptional group
E$_6$ \cite{R3}, which is suggested as a low energy limit of superstring
theories. In this group each generation of fermions lies in the
representation of dimension 27, and in addition to the usual SM chiral fields,
twelve new fields are needed to complete the representation. Among these, there
will be two weak isodoublets of heavy leptons, a right--handed and a
left--handed one. \sx

\nn $iii)$ Mirror fermions: they have chiral properties which are opposite to
those of ordinary fermions, i.e. the right--handed components are in weak
isodoublets and the left--handed ones are in weak isosinglets; there is also a
left--handed heavy neutrino \cite{R4}. These fermions appear in many extensions
of the SM and provide a possible way to restore left--right symmetry at the
scale of the electroweak symmetry breaking; they naturally occur in lattice
gauge theories. \sx

\nn $i$v) Singlet fermions: these are the most discussed fermions in the
literature, a prominent example being the SO(10) neutrino \cite{R5,R6}. Indeed,
in this unifying group, which is one of the simplest and most economic
extensions of the SM, the smallest anomaly--free fermion representation has
dimension 16. It contains a right--handed neutrino in addition to the 15 Weyl
fermions in one fermion generation; this neutrino is of the Majorana type.
Singlet neutrinos, which can be either of Majorana or Dirac type, and quarks
also occur in E$_6$ \cite{R3}. \\

\nn The search for these new fermions will be a major goal of the next
generation of accelerators. In particular, high--energy $\ee$ colliders
provide unique facilities to search for these fermions, and in case of
discovery, to explore their basic properties. \s

\nn In a preceding paper \cite{moi}, hereafter referred to as (I), one of us
discussed in a rather general context, the production mechanisms in $\ee$
collisions and the decay modes of the new fermions. Here, we will complete this
discussion by analyzing the discovery potential of the next $\ee$ linear
collider. Concentrating on new leptons, which are difficult to search for at
hadron colliders \cite{R7}, we study in great details their possible signatures
and the corresponding backgrounds at a future high--energy $\ee$ collider with
a center of mass energy of 500 GeV. To be as close as possible to the
experimental conditions, we use a model detector for $\ee$ collisions at this
energy, exploiting the various experimental studies which have been carried out
on the potential of a 500 GeV $\ee$ machine \cite{EE500}. \\

\nn The paper is organized as follows. In the next section, we first introduce
the interactions of the new fermions, discuss their decay modes and summarize
the existing constraints on their masses and couplings. In section 3, we study
the pair production of neutral and charged heavy leptons. In section 4, we
analyze in detail the signals and backgrounds for the single production of
heavy leptons in association with their ordinary partners, paying a special
attention to the first generation where additional production channels are
present. Finally, section 5 contains our conclusions.

\subsection*{2.~Physical Basis}

\renewcommand{\theequation}{2.\arabic{equation}}
\setcounter{equation}{0}

In this section, we briefly  summarize the couplings of the new fermions and
their dominant decay modes, concentrating on those features which will be
needed in the present analysis. A more general discussion can be found in (I).
\\

\nn {\bf 2.1~Interactions} \\

\nn Except for singlet neutrinos which have no electromagnetic and weak
charges, the new fermions couple to the photon and/or to the electroweak gauge
bosons $W/Z$ with full strength. These couplings allow for the pair production
in $\ee$ annihilation of heavy leptons and quarks, if their masses are smaller
than the beam energy. To calculate the production cross sections, one needs the
vector and axial--vector couplings of the new fermions to the photon and $Z$
boson; in units of the proton charge, they are given by
\beq
v^F_\gamma = e^F \ \ , \ \ \ a^F_\gamma=0 \ \ , \ \ \
v^F_Z \equiv v_F = \frac{2I_{3L}^F+2I_{3R}^F- 4e^F s_W^2}{4s_W c_W}
\ \ , \ \ \ a^F_Z \equiv a_F = \frac{2I_{3L}^F-2I_{3R}^F}{4s_W c_W}
\eeq
where $e^F$ is the electric charge of the fermion $F$, $I^{F}_{3L}, I_{3R}^F$
the third components of the left--handed and right--handed weak isospin and
$s_W^2=1-c_W^2 \equiv\sin^2 \theta_W$. This expression clearly exhibits the
facts that vector fermions have no axial--vector couplings, and that the
axial--vector couplings of sequential and mirror fermions are of opposite
signs. \s

\nn If they have non--conventional quantum numbers, the new leptons and quarks
will mix with the SM ordinary fermions which have the same U(1)$_{\rm Q}$ and
SU(3)$_{\rm C}$ assignments. This mixing will give rise to new currents which
determine to a large extent their decay properties and allow for a new
production mechanism: single production in association with their light
partners; see Fig.~1. The mixing pattern depends sensitively on the considered
model and, in general, is rather complicated especially if one includes the
mixing between different generations. However, this inter--generational mixing
should be very small since it would induce at the tree level, flavor changing
neutral currents which are severely constrained by existing data \cite{R3}. \s

\nn In the present analysis, we will neglect the inter--generational mixing and
treat the few remaining mixing angles as phenomenological parameters. To
describe our parametrization, let us explicitely write down the interaction of
the electron and its associated neutrino with exotic charged and neutral heavy
leptons. Allowing for both left--handed and right--handed mixing, and assuming
small angles so that one can write $\sin\zeta_{L,R} \simeq \zeta_{L,R}$, the
Lagrangian describing the transitions between $e,\nu_e$ and the heavy leptons
$N,E$ of the first generation is
\begin{eqnarray}
{\cal L} &=& g_W \left[ \zeta_L^{\nu E} \bar{\nu_e}
\gamma_\mu E_L + \zeta_R^{\nu E} \bar{\nu_e} \gamma_\mu E_R \right] W^\mu
+ g_Z \left[ \zeta_L^{eE} \bar{e} \gamma_\mu E_L
+ \zeta_R^{eE} \bar{e} \gamma_\mu E_R \right] Z^\mu \ + \ {\rm h.c.} \non \\
&+& g_W \left[ \zeta_L^{eN} \bar{e}
\gamma_\mu N_L + \zeta_R^{eN} \bar{e} \gamma_\mu N_R \right] W^\mu
+ g_Z \left[ \zeta_L^{\nu N} \bar{\nu_e} \gamma_\mu N_L
+ \zeta_R^{\nu N} \bar{\nu} \gamma_\mu N_R \right] Z^\mu \ + \ {\rm h.c.}
\hspace*{0.5cm}
\end{eqnarray}
\nn with $g_W=e/\sqrt{2}s_W$ and $g_Z= e/2s_Wc_W$. The generalization to the
other lepton families and to quarks is obvious. Note that in principle, the
indices {\small L,R} refer to the handedness of the heavy lepton mixing with
its light partners, but since the latter are almost massless they are also the
chirality of the heavy leptons. \\

\nn {\bf 2.2~Decay Modes} \\

\nn The heavy leptons decay through mixing into massive gauge bosons plus their
ordinary light partners; for masses larger than $M_W/M_Z$ the vector bosons
will be on--shell and decay into light quarks and leptons; Fig.~1c. Using the
scaled masses $v_{W,Z}=M_{W,Z}^2/m_F^2$, the partial decay widths are
\begin{eqnarray}
\Gamma (F_{L,R} \ra Zf) &=& \frac{\alpha} {32 s_W^2 c_W^2} \left(
\zeta_{L,R}^{fF} \right)^2 \ \frac{m_F^3}{M_Z^2}  ( 1-v_Z)^2 (1+2v_Z) \non \\
\Gamma (F_{L,R} \ra Wf') &=& \frac{\alpha}{ 16 s_W^2 c_W^2} \left(
\zeta_{L,R}^{f'F} \right)^2 \ \frac{m_F^3}{M_Z^2}  ( 1-v_W)^2 (1+2v_W)
\end{eqnarray}
\nn For small mixing angles, the heavy fermions have very narrow widths: for
$\zeta_L/\zeta_R \sim 0.1$ and masses around 100 GeV the partial decay widths
are less than 10 MeV. The decay widths increase rapidly for increasing fermion
masses, $\Gamma \sim m_F^3$~, but for allowed values of the mixing angles [see
below], do not exceed the 100 GeV range even for $m_F \sim {\cal O} (1$~TeV).
The charged current decay mode is always dominant and for fermion masses much
larger than $M_Z$, the branching ratios are 1/3 and 2/3 for the neutral and
charged current decays, respectively. Note that for Majorana neutrinos,  both
the $l^-W^+ /\nu_l Z$ and $l^+ W^- / \bar{\nu_l} Z$ are possible; this makes
its total decay width twice as large as for Dirac neutrinos. However, for small
mixing angles and/or moderate masses, the widths are too small to be resolved
experimentally and no distinction between Dirac and Majorana neutrinos can be
made at this level. \s

\nn To fully reconstruct the heavy fermion from its final decay products one
needs the branching ratio of its decay into visible particles. Neglecting the
mixing between different generations, the branching ratios for the decays of
the heavy leptons into charged ordinary leptons and gauge bosons which
subsequently decay into jets take the asymptotic values [for large $m_L$]
\begin{eqnarray}
{\rm Br}(E^- \ra Z \ l^- \ra jj l^-) &\simeq & \frac{1}{3} \times 0.70 \
\simeq \ 0.23 \non \\
{\rm Br}(N \ra W^+ l^- \ra jj l^-) & \simeq & \frac{2}{3} \times 0.66 \
\simeq \ 0.43
\end{eqnarray}

\nn In the case of $E$, one can also include the cleaner $Z\ra \ee +\mu^+\mu^-$
decays, but the branching ratio is rather small: $\sim 6\%$ compared to $\sim
70\%$ for $Z \ra$ hadrons. \s

\nn For fermions which belong to the same isodoublets, the decay of the
heaviest fermion into the lighter one plus a $W$ boson is in principle also
possible. However, since the new fermions must be approximately degenerate not
to induce large radiative corrections \cite{R2} to electroweak parameters such
as $M_W$ and $\sin^2\theta_W$, the $W$ boson should be off--shell and these
decays are strongly suppressed unless the mixing angles are prohibitively
small.
These cascade decays, the amplitudes of which can be found in (I), will not be
considered here. \s

\nn Finally, if the new fermions have masses smaller than $M_W$, the
intermediate vector bosons will be off--shell and the decays are three--body
decays. Complete expressions for partial widths, angular and energy
distributions can be found in (I). In this paper, we will assume that the new
fermions are always heavier than the $W$ boson; this will be justified below.\\

\nn {\bf 2.3.~Constraints} \\

\nn Let us now briefly summarize the present experimental constraints on
the masses of the new fermions and on their mixing with the ordinary fermions.
The mixing will alter the couplings of the electroweak gauge bosons to
light quarks and leptons from their SM values. Since the couplings
of the latter to the $Z$ boson have been very accurately determined at LEP100
[through the measurement of total, partial and invisible decay widths as well
as forward--backward and polarization asymmetries] and found to agree with the
SM predictions up to the level of one percent, the mixing angles
are constrained to be smaller than ${\cal O}(10^{-1})$ \cite{R8}.
In the case of leptons, if the left and right--handed mixing angles have
the same size, the precise measurement of the anomalous magnetic moments of the
electron and muon leads to even more stringent constraints $\theta_{\rm mix}<
{\cal O}(10^{-2})$ \cite{R9}. Indeed, without the chiral protection $\zeta_L$
or $\zeta_R \sim 0$, the contribution of heavy lepton loops to $(g-2)_{e,\mu}$
will be proportional to $m_e/m_{L}$ [rather than to $m_{e,\mu}^2/m_{L}^2$] and
very small $\theta_{\rm mix}$ and/or extremely large $m_L$ are needed to
protect the electron and muon from acquiring a too large magnetic moment. \s

\nn From the negative search of new states and from the measurement of $Z$
decay widths at LEP100, one can infer a bound of the order of $M_Z/2$ on the
masses of the new fermions \cite{R10} independently of their mixing, except for
singlet neutrinos which have no full weak couplings to the $Z$ boson.
Masses up to $m_F \sim M_W$ can be probed at LEP200. \s

\nn In the case of heavy neutrinos, including the gauge singlets, an additional
constraint is provided by the negative search \cite{R11,R11a} of these states
through single production in $Z$ decays: $Z \ra \bar{\nu}_e +N \ra \bar{\nu_e}+
e^-W  \ra \bar{\nu}_e + e^-jj$. If the $\nu N$ mixing angle is of the order of
$\sim 0.1$ or larger, the neutrino must be heavier than the $W$ boson
\cite{R11}.  A similar mass bound can be established for the charged lepton,
which in $Z$ decays leads to the same final state, $Z \ra e^+E^- \ra e^+ \nu_e
W^- \ra \nu_e e^- jj$, with approximately the same rate. Note that for mixing
angles much smaller than ${\cal O}(10^{-2})$, no bound can be derived on the
singlet neutrinos masses: the production cross section is small and/or the
heavy neutrino escapes detection because of its too large decay length. \s

\nn Since for quarks and for the new leptons which have full couplings to the
$Z$ boson, masses smaller than $M_W$ can be probed at LEP200 in pair
production,
and because the singlet neutrinos with such masses and with not too small
mixing angles [which would prohibit their single production also at higher
energies] are already ruled out, we will assume in our discussion at a 500 GeV
collider, that the new fermions are heavier than the $W$ boson. \s

\nn Finally, let us mention the constraints on the new gauge bosons which are
also predicted by extended gauge models and which might have observable effects
in the production and decay of the new fermions. Direct searches at present
hadron colliders and indirect searches through the high--precision LEP
measurements lead to bounds of the order of 200--400 GeV on the new vector
boson mass, depending on the particular chosen model \cite{R2}; the accessible
mass range can be pushed up to $ \sim 500$ GeV in the near future. Masses up to
3 TeV can be probed at the 500 GeV collider itself \cite{nous}. In our
analysis,
we will assume that these new gauge bosons are heavy enough not to affect the
physics at the 500 GeV scale. The mixing between these new gauge bosons and the
standard $Z$ boson would alter the couplings of the latter to ordinary and new
fermions. However, since it is constrained to be smaller than ${\cal
O}(10^{-2})
$ \cite{R2}, this mixing will also be neglected in the present analysis.

\subsection*{3.~Pair production}

\renewcommand{\theequation}{3.\arabic{equation}}
\setcounter{equation}{0}

\vspace*{3mm}

\nn {\bf 3.1~Cross sections and distributions} \\

\nn If their masses are smaller than the beam energy, the new fermions can be
pair produced in $\ee$ collisions; Fig.~1a. For charged fermions the process
proceeds through $s$--channel $\gamma$ and $Z$ boson exchange, while for
non--singlet heavy neutrinos only $Z$ boson exchange is present. There are also
contributions from $t$--channel $W/Z$ exchange in the case of the first
generation of heavy leptons, but they are quadratically suppressed by mixing
angle factors and therefore totally negligible for experimentally allowed
values of these angles.  \s

\nn The differential cross section ${\rm d}\sigma/{\rm d}\cos \theta$ [with
$\theta$ specifying the	direction of the fermion $F$ with respect to the
incoming electron] for the process $\ee \ra F \bar{F}$ has been given in (I)
in the general case where several channels are present. In the case where only
$\gamma$ and $Z$ $s$--channels exchanges contribute, the expression simplifies
to
\begin{eqnarray}
\frac{{\rm d} \sigma}{{\rm d} \cos \theta } = \frac{3}{8} \sigma_0N_c \beta_F
\left[ (1+ \cos^2 \theta) (\sigma_{VV} + \beta^2_F \sigma_{AA}) +(1-\beta^2)
\sin^2\theta\sigma_{VV} +2 \beta_F \cos \theta \sigma_{VA}\right]
\end{eqnarray}

\nn where $N_{c}$ is a color factor, $\sigma_0=4\pi \alpha^2/3s$ the
point--like QED cross section for muon pair production and $\beta_F= (1-4m_F^2/
s)^{1/2}$ is the velocity of the fermion in the final state; in terms of
the $FF \gamma$ and $FFZ$ couplings, the reduced cross sections read
[neglecting the small decay width of the $Z$ boson]

\beq
\sigma_{VV} &=& e_e^2 e_F^2 + \frac{2 e_ee_F v_e v_F}{1-M_Z^2/s} + \frac{
(a_e^2+ v_e^2)v_F^2}{(1-M_Z^2/s)^2} \non \\
\sigma_{AA} &=& \frac{(a_e^2+ v_e^2)a_F^2}{(1-M_Z^2/s)^2} \non \\
\sigma_{VA} &=& \frac{2 e_e e_F a_e a_F}{1-M_Z^2/s} + \frac{4v_ea_e v_F a_F}
{(1-M_Z^2/s)^2}
\eeq
\nn The total production cross sections are simply given by
\beq
\sigma_F = \sigma_0 N_c \ \left[ \frac{1}{2} \beta_F (3-\beta_F^2) \sigma_{VV}
+ \beta_F^3 \sigma_{AA} \right]
\eeq
\nn The cross sections for the pair production of mirror and vector heavy
leptons are displayed in Fig.~2 for a center of mass energy of 500 GeV. For
charged leptons, they are of the order of 0.5 pb for masses not too close to
the beam energy. With the expected integrated luminosity of 20 fb$^{-1}$ for a
500 GeV $\ee$ collider, this corresponds to $\sim $ 10,000 events. Even for
masses very close to the kinematical limit, $\sim$ a few GeV, the cross
sections exceed the 100 fb level, leading to a large number of events. Due to
the absence of the photon exchange, the cross sections for the heavy neutrinos
are smaller than for charged leptons. But the event rates are also very large,
especially for vector neutrinos, allowing for the discovery of these leptons up
to the kinematical limit. For instance, in the case of the mirror neutrino
[which has the smallest cross section] with a mass of 247.5 GeV, one is still
left with $\sim 200$ events a year, asumming $\int {\cal L} = 20~{\rm
fb}^{-1}$. \s

\nn The angular distributions are shown in Fig.~3 for lepton masses of 150 GeV.
Because they have no axial--vector couplings, vector leptons have
forward--backward symmetric angular distributions [the term proportional to
$\cos\theta$ is zero] contrary to mirror fermions for which d$\sigma$/dcos
$\theta$ is larger in the backward--direction. The forward--backward asymetries
are given by
\beq
A_F = \frac{3}{4} \beta_F \frac{\sigma_{VA}}{\frac{1}{2}(3-\beta^2_F)
\sigma_{VV}+ \beta^2_F \sigma_{AA}}
\eeq
For mirror leptons the asymmetries are sizeable [except near threshold where
they vanish] and negative, contrary to the case of sequential heavy leptons for
which they have the same magnitude but with the opposite sign. Of course, for
vector leptons, the forward--backward asymmetries are zero. \s

\nn The	polarization four--vector $P_\mu$ of the final state fermion $F$ is
defined by d$\sigma^{{\rm pol}}(\cos \theta)=$ d$\sigma^{{\rm unpol}}(\cos
\theta) \times [1+ P_\mu n^\mu]$, with $n_\mu$ the spin vector.  In the $F$
rest frame, the components are $(0, P_\perp, 0, P_{||})$ with $P_\perp $ and
$P_{||}$ being the transverse and longitudinal polarizations with respect to
the flight direction; their expressions can be found in (I). The two components
are shown in Fig.~4 for lepton masses of 150 GeV, as a function of the
scattering angle. For vector leptons, the magnitude of $P_\perp$ and $P_{||}$
is four times larger for $E$ than for $N$ but [due to the fact that their
couplings to both the photon and the $Z$ are vectorial] the shape is exactly
the same; $P_\perp$ is positive, forward--backward  symmetric and maximal
(minimal) for $\cos \theta=\pm1 (0)$ while $P_{||}$ is monotonically decreasing
for increasing $\cos \theta$ and is zero for $ \theta=\pi/2$. Mirror charged
leptons have very small transverse and longitudinal polarizations contrary
to their neutral partners. \s

\nn Averaged over the polar angle, the longitudinal polarization $<P_{||}>$ is
zero for vector leptons, while $P_\perp$ is small and positive. For mirror
leptons, the two components have the same absolute values as for sequential
leptons, only $P_{||}$ has the opposite sign. \s

\nn Hence, the final polarization and the angular distributions of the produced
leptons are very usefull to discriminate between different types of particles:
mirror, vector and also sequential fermions. \\

\nn {\bf 3.2.~Signals and backgrounds} \\

\nn In this subsection, we discuss the signals and backgrounds for the pair
production of the new fermions, concentrating on the heavy leptons which are
more difficult to search for at hadron colliders \cite{R7}. \s

\nn The pair production of charged and neutral heavy leptons leads to final
states with two ordinary leptons and two massive gauge bosons which
subsequently decay into four massless fermions. To fully reconstruct the heavy
leptons from their decay products, one needs the final states with charged
ordinary leptons from the first vertex and jets from the decay of the gauge
bosons [in the case of the $Z$ boson one can also include the $Z \ra \ee +
\mu^+ \mu^-$ decays, but the branching ratio is rather small, $\sim 6\%$
compared to $\sim 70\%$ for $Z \ra$~hadrons]. Using the notation of the first
generation, the branching fractions for these final states for large lepton
masses are
\begin{eqnarray}
\ee \lra E^+ E^- & \lra & e^+e^- ZZ  \ \ \ \ \lra \ \ e^+ e^- + 4~{\rm jets}
\hspace{1.5cm} {\rm B.R.} \simeq 5.5 \% \non \\
\ee \lra \bar{N} N & \lra & e^+e^- WW \ \ \lra \ \ e^+ e^-  + 4~{\rm jets}
\hspace{1.5cm} {\rm B.R.} \simeq 20 \%
\end{eqnarray}

\nn The branching fraction for this specific final state is four times larger
in the case of neutral leptons than for charged leptons but since the
production cross sections for the latter are a factor 2.5 to 5 larger than for
heavy neutrinos, the number of events for the two processes are of the same
order. For instance, for a  heavy lepton with a mass of 150 GeV, assuming a
luminosity $\int {\cal L}=20~ {\rm fb}^{-1}$, one has 520 and 460 events for
vector and mirror electrons respectively, while for vector and mirror neutrinos
one has 820 and 320 events respectively. \s

\nn Several conventional processes lead to the same final states, the most
important ones being triple gauge boson production:
\begin{eqnarray}
\ee \lra \ \ ZZZ [\gamma ] & \lra & (e^+e^- ) \ + \ (jj)(jj) \non \\
\ee \lra WWZ [\gamma ] & \lra & (e^+e^- ) \ + \ (jj)(jj)
\end{eqnarray}
\nn However, compared to the signals, these processes are of higher order
[suppressed by one extra power of $\alpha$] and the cross sections are rather
small: if one neglects the Higgs induced contributions, one has $\sigma(\ee \ra
WWZ) \simeq$ 40 fb and $\sigma(\ee \ra ZZZ)$ is only $\simeq $ 1 fb. If in
addition the $Z$ boson is forced to decay into charged leptons, the cross
sections drop by more than an order of magnitude. For a luminosity
of 20 fb$^{-1}$ one has only 25 $\ee+4~{\rm jet}$ events from the dominant
$WWZ$ process, to be compared with more than 300 events from the signals for
$m_L \simeq 150$ GeV. The cross sections for $\ee \ra ZZ \gamma $ and $WW
\gamma$ are extremely small after forcing the virtual photon to have a mass of
the order of $M_Z$. Other background processes, lead to negligible cross
sections at a c.m. energy of 500 GeV. \s

\nn To be as close as possible to the experimental conditions, we will use in
the following, a model detector for $\ee$ collisions at a c.m. energy of 500
GeV
to simulate the efficiencies for reconstructing the signal events. For
illustration, we will study the case of the pair production of charged
leptons [the cross sections for mirror and vector leptons are nearly equal]
with a mass of 200 GeV; before any cut is applied the cross section is
400 fb leading to 450 $\ee + 4~{\rm jet}$ events for an integrated luminosity
of 20 fb$^{-1}$. Since the cross sections for the production of the other types
of leptons are approximately the same, the results for neutral heavy leptons
should be similar. The background reactions will not be simulated since the
cross sections are negligible, even before one requires that the invariant mass
of one lepton and two jets combine to give the mass of the heavy lepton. \\

\nn The process of pair production was simulated in PYTHIA5.6 \cite{Pythia} by
allowing for a fourth generation of fermions, which was then made to correspond
to an excited copy of the first generation. Only the decay channel $E
\rightarrow Z e$ was turned on, with a weak matrix element similar to the one
of the standard $l \rightarrow W \nu$ process. This simulation should be
representative, in terms of detector effects, of pair production of any lepton
type. \s

\nn In all cases, full hadronization was allowed to take place.
Detector effects were taken into account by subjecting the particles to
resolution smearing and acceptance cuts, as summarized in Table 1. The
parameters for the model detector are similar to  the ``standard'' set of
Ref.~\cite{GrosseWiesmann},  but with angular acceptance up to $|\cos(\theta)|
< 0.98$ for the electromagnetic calorimeter as well as for the charged particle
tracker. This ``standard" detector has similar properties to the ones of
present high--energy colliders: in this respect we took a rather conservative
approach since some improvements for the next generation detectors are
expected. \s

\nn All charged particles had their momentum ``measured'' by the central
tracker, whereas the energy of photons and electrons was ``reconstructed'' with
the resolution smearing appropriate for the electromagnetic calorimeter. In the
case of electrons, the combined measurement of momentum and energy served to
define the ``reconstructed'' 4--vector. Muons were assumed to be identifiable
by
means of hypothetical muon chambers, or other subdetector type. Finally, other
particles were assumed to be detected by the hadronic calorimeter. \\

\nn As explained above, the simulation of detector effects imposed minimum
requirements on the particle directions and momenta. The loss in solid angle is
small in this case [but not for processes with an angular distribution peaked
along the beam line]. A further loss in efficiency resulted from the lepton
selection criterion: in order to minimize backgrounds, only electrons with
momentum higher than $p_l^{\rm min} = 30$ GeV were considered. For pair
production of not--too--heavy leptons, $p_l^{\rm min}$ should be set lower. \s

\nn In order to reconstruct the event consisting of four jets and two leptons,
the following procedure was adopted:
\begin{description}
\item[(i)] \ \  Accept only events having one and only one ``detected''
lepton pair.
\item[(ii)] \ Excluding these two lepton tracks, construct four jets out of the
remaining particles, using the Durham algorithm \cite{Durham}.
\item[(iii)] Associate a pair of jets with each lepton in such a way that the
following $\chi^2$ is minimized: $\chi^2 = (E_{\rm beam} -E_{j1} -E_{j2}
-E_{e^+
} )^2 (E_{\rm beam}-E_{j3} -E_{j4} -E_{e^-} )^2$.
\end{description}

\nn Fig.~5 shows the resulting mass histogram of one of the reconstructed $E^+$
for the case of a pair of charged heavy leptons of mass 200 GeV [an exactly
analogous histogram can be made for $E^-$]. The tails of the distribution can
be cleanly removed, as shown by the shaded area in the figure, with a loss of
efficiency of 40\%, by requiring that each of the invariant jet--jet masses be
consistent with the mass of a $Z$ boson within 15 GeV. Note that only 6\% of
the
events did not pass the cuts on the solid angle and the minimum lepton energy.
The histogram is normalized for an integrated luminosity of 20 fb$^{-1}$ and in
this case, one is left with 250 events [in the dashed area] after all cuts were
applied. \s

\nn As discussed above, similar numbers can be obtained for the other lepton
types. Therefore, it is clear that the detection of pair--produced heavy
leptons
should be very easy at $\ee$ colliders.

\subsection*{4.~Single production}

\renewcommand{\theequation}{4.\arabic{equation}}
\setcounter{equation}{0}

\vspace*{2mm}

\nn {\bf 4.1.~Cross sections and distributions} \\

\nn In $\ee$ collisions one can also have access to the new fermions via single
production in association with light fermions, if fermion mixing is not too
small; Fig.~1b. The process proceeds through $s$--channel $Z$ exchange in the
case of quarks and second/third generation leptons. For the first generation
heavy leptons, one has to include additional $t$--channel gauge boson
exchanges:
$W$ exchange for $N$ and $Z$ exchange for $E$. Complete expressions for the
angular distributions and the total cross sections in the general case can be
found in (I). Neglecting the mass of the light fermion partner, the total cross
sections for leptons of the second and third generations and for quarks take
the simple form

\begin{eqnarray}
\sigma(F_{L,R}) = \sigma_0 N_c \frac{ \left(\zeta_{L,R}^{Ff} \right)^2 }
{128 c_W^4s_W^4} (a_e^2 + v_e^2) \ \frac{(1-\mu^2)^2(1+\mu^2/2)}{(1-z)^2}
\end{eqnarray}
where now $a_e=-1,~v_e=-1+4s_W^2$, $\mu=m_F^2/s$ and $z=M_Z^2/s$. The
expressions of the cross section for the production of the first generation
heavy leptons are more involved in the general case. However, for only $Z$ and
$W$ exchange, once all the couplings are specified, they take the rather simple
form
\begin{eqnarray}
\sigma(E_{L,R}) & = & \sigma_0 \frac{3(\zeta_{L,R}^{eE} )^2}{256c_W^4s_W^4}
\left\{ (v_e \pm a_e )^2 \left[ - \left( 2-\mu^2+2z+2 (1+z-\mu^2)\frac{1+z}
{1-z} \right) L_Z \right. \right. \non \\
&+& \left. \frac{1}{3} \frac{(1-\mu^2)^2(1+\mu^2/2)}
{(1-z)^2} + \frac{(1-\mu^2)(3+2z-\mu^2)}{1-z} +\frac{(1-\mu^2)(1+2z)}{z}
\right] \non \\
&+& (v_e \mp a_e)^2 \left[ - \left((2z -\mu^2)(1-z)+2z (z-\mu^2) \right)
\frac{1}{1-z} L_Z \right. \\
&+& \left. \left. \frac{1}{3} \frac{(1-\mu^2)^2(1+\mu^2/2)}
{(1-z)^2} -\frac{(1-\mu^2)(1-2z+ \mu^2)}{1-z}+ \frac{(1-\mu^2)(1+2z-2\mu^2)}
{1+z-\mu^2} \right] \right\} \non
\end{eqnarray}
\begin{eqnarray}
\sigma(N_R) &=& \frac{3\sigma_0}{16 s_W^4} \left\{\frac{1}{24} \frac{(\zeta_R^{
N\nu})^2}{c_W^4} (v_e^2+a_e^2) \frac{(1-\mu^2)^2(1+\mu^2/2)}{(1-z)^2} \right.
\non \\
&-& \frac{v_e+a_e}{4c_W^2} \frac{ \zeta_R^{Ne} \zeta_R^{N\nu}}{1-z}
\left[ (1-\mu^2)(1-2w+\mu^2)+2w(w-\mu^2) L_W \right] \non \\
&+& \left. (\zeta_R^{Ne})^2 \left[ \frac{(1-\mu^2)(1+2w- 2\mu^2)}
{1+w-\mu^2} + (\mu^2-2w) L_W \right] \right\}
\end{eqnarray}
\begin{eqnarray}
\sigma(N_L) &=& \frac{3\sigma_0}{16s_W^4} \left\{ \frac{1}{24}
\frac{(\zeta_L^{N\nu})^2}{c_W^4} (v_e^2+a_e^2) \frac{(1-\mu^2)^2(1+\mu^2/2)}
{(1-z)^2} \right. \non \\
&+& \frac{v_e+a_e}{4c_W^2}\frac{\zeta_L^{N\nu} \zeta_L^{Ne}}{1-z}
\left[ (1-\mu^2)(3+2w-\mu^2)-2(1+w-\mu^2)(1+w) L_W \right] \non \\
&+& \left. (\zeta_L^{Ne})^2 \left[ \frac{(1-\mu^2)(1+2w)}{w} -
(2-\mu^2+2w) L_W \right] \right\}
\end{eqnarray}
where for $V=W,Z$ one has
\beq
L_V= \log \frac{1-\mu^2+v}{v} \hspace*{1cm} {\rm with} \ \ v=M_W^2/s\ \ {\rm
or}
\ \ M_Z^2/s
\eeq
\nn The cross sections are the same for the charge conjugate states. \s

\nn The cross section for the single production of the heavy leptons in
association with their light partners at a 500 GeV c.m. energy are displayed in
Fig.~6 as a function of $m_L$; for the left and right--handed mixing angles, we
have assumed values close to the experimental bounds $\zeta_{L,R}^{Ne}=\zeta_{
L,R}^{N\nu} =0.1$. In the case of $N_L$ production, the cross section is of the
order of 1 pb for masses of ${\cal O}$(100 GeV), which corresponds to $2 \times
10^{4}$ events per year with a luminosity of 20 fb$^{-1}$. It drops to $\sim
0.1$ pb for $m_N \sim 450$ GeV. For smaller $\zeta_L$ values the cross section
has to be scaled down; for $\zeta_L$ values ten times smaller, one is left with
$\sim $1 fb for $m_N \simeq$ 450 GeV. For $N_R$ production, the cross section
is more than one order of magnitude smaller than $\sigma(N_L)$ for the same
value of the mixing angle. It is practically constant for most of the relevant
range of $m_N$, being approximately 40 fb, and it starts to drop only for
$m_N\sim 450$ GeV. For this mass value and for $\zeta_R =0.01$, one is left
with only $\sim 10$ events for an integrated  luminosity of 20 fb$^{-1}$. \s

\nn The cross sections for the production of $\bar{E}e$ and $E\bar{e}$ pairs
through right and left--handed mixing, which are approximately equal due to the
fact that the vectorial coupling of the electron to the $Z$ is practically zero
for $s_W^2 \simeq 1/4$, are one order of magnitude smaller than in the case of
$N_L$. Nevertheless, $\sigma(E)$ is large enough for the value chosen  for the
mixing angles, to allow for the possibility of a large sample of events: for
$m_E \simeq 400$, and a mixing angle of $\zeta \simeq 0.02$, one has 20 events
for the expected luminosity of the $\ee$ collider. \s

\nn The largest contribution to the production cross sections of these heavy
leptons comes from the $t$--channel $Z$ or $W$ exchanges. This can be seen by
comparing these cross sections with those of the second and third generation
leptons where the process is mediated only by $s$--channel $Z$ exchange. Since
the latter cross sections are less than a few fb, i.e. three orders of
magnitude
smaller than $\sigma(N_L)$, it seems rather unlikely that second and third
generation neutrinos and charged leptons can be singly produced with sufficient
rates. We will therefore concentrate on the first generation in what follows.\s

\nn The differential cross sections are shown in Fig.~7. In the case of Dirac
$N_L$ production, d$\sigma/$d$ \cos \theta$ is peaked in the forward direction,
while for $N_R$ it is practically flat except near $\cos \theta$=1. For the
production of Majorana particles, the angular distribution is forward--backward
symmetric and in the case of $N_L$, it is peaked in both directions \cite{R6}.
Exploiting the different  behavior of d$\sigma/$d$ \cos \theta$, it should be
rather easy to distinguish between left--handed and right--handed mixings and
between Dirac and Majorana neutrinos. As in the case of $N_L$, the angular
distributions for $E$ production d$\sigma/$d$ \cos \theta$ are peaked in the
forward direction due the $t$--channel $Z$ exchange. Since they are practically
the same for $E_L$ and $E_R$ production, it is very difficult to distinguish
between left and right--handed mixing. In this respect, the polarization of the
final heavy state can be very useful. \\

\nn The longitudinal and transverse components of the  polarization vector of
heavy fermions produced in association with light partners have been given in
(I). In Fig.~8, the two components of the polarization vectors of $E$ and $N$
are shown as a function of the scattering angle and for a lepton mass of 400
GeV. In the case of $N$, the transverse components are positive and similar in
shape for left--handed and right--handed mixing; the longitudinal polarizations
are rather different for $N_L$ and $N_R$. In the case of the charged leptons,
both longitudinal and transverse components are completely different for left
and right--handed mixings, in particular, they are of opposite signs for a
given value of $\theta$. This important feature will be useful to distinguish
$E$ particles which couple to the $Z$ boson with left or right--handed
currents, in particular as the distinction is rather difficult using the
angular distributions. \\

\nn {\bf 4.2.~Signals and backgrounds} \\

\nn The single production of the heavy leptons leads to less complicated final
states than for pair production: just two ordinary leptons, one coming from the
production process and the other from the decay of the heavy lepton, and a
gauge
boson which subsequently decays into two massless fermions. As previously
discussed for the case of pair production, to fully reconstruct the heavy
leptons from their decay products one has to concentrate on the following final
states
\begin{eqnarray}
\ee \lra E^\pm e^\mp & \lra & e^\pm e^\mp Z  \ \ \ \ \lra \ \ e^+ e^- + 2~{\rm
jets} \hspace{1.5cm} {\rm B.R.} \simeq 23 \% \non \\
\ee \lra \bar{\nu} N/ \nu \bar{N} & \lra & \nu_e e^\pm  W^\mp  \ \ \lra \ \
\nu_e e^\pm + 2~{\rm jets} \hspace{1.5cm} {\rm B.R.} \simeq 43 \%
\end{eqnarray}
where in the case of $N$, $\nu_e$ stands for the electronic neutrino and its
conjugate state. The branching fractions are for large lepton masses compared
to $M_W$ and for $E$ one can also use the decays $Z \ra \ee/\mu^+ \mu^-$. \s

\nn In the following we will simulate the single production of the heavy
leptons which leads to the final states shown above, as well as the
corresponding background reactions. We will concentrate on the case of
the first generation neutral and charged leptons and assume a left--handed
mixing with $\theta_{\rm mix}$ taken to be $\theta_{\rm mix}=0.025$ for $N$
and $0.05$ for $E$. \s

\nn The processes for the production of the charged and neutral heavy leptons
$E$ and $N$ were simulated by incorporating in the PYTHIA \cite{Pythia}
generator the matrix elements for the three body reactions: $ e^+ e^- \ra N
\nu_e / E^\pm e^\mp \ra \nu_e e^\pm W^\mp /e^\pm e^\mp Z$ and forcing the
gauge bosons to decay hadronically; full hadronization was allowed to take
place. All the resulting particles were then subjected to detector resolution
and acceptance effects. The parameters for the detector are shown in Table 1.
In the case of the neutral heavy lepton, the missing momentum vector was
calculated and subsequently assumed to be the reconstructed neutrino momentum.
The background processes were simulated using existing parameter options in
PYTHIA. \s

\nn In the case of the charged lepton $E$, the signal consists of an $\ee$ pair
and two jets. Other processes likely to produce such a configuration are:
\begin{description}
\item[(i)] \ \ $ e^+ e^- \to e^+ e^- Z$, with the $Z$ decaying hadronically;
the cross section is 3800 fb.
\item[(ii)] \ $ e^+ e^- \to Z Z$, where the cross section is 615 fb after
forcing
one of the $Z$ bosons to decay hadronically and the other into electrons.
\item[(iii)] $ e^+ e^- \to t \bar{t} $, followed by  $ t \to b W$ and leading
to two electrons and 2 jets but with missing momentum.
\item[(iv)] $ \gamma \gamma \to e^+ e^- q \bar{q}$ which has a large cross
section but the jets have small invariant masses and the resulting  events have
the primary electrons going mostly along the beam pipe.
\end{description}

\nn In the case of the neutral lepton $N$, the signal consists of an electron,
a pair of jets and missing momentum. The backgrounds that one has to consider
are:
\begin{description}
\item[(i)] \ \ $ \ee   \to e \nu W$ with a cross section of 5800 fb when the
$W$ boson decays hadronically.
\item[(ii)] \ $ \ee \to WW$ where one of the $W$'s decays hadronically and the
other to an $e \nu$ pair and the cross section in this case is 1140 fb;
\item[(iii)] $ \gamma \gamma \to e (e)  q \bar{q} $ where one of the electrons
escapes observation; the cross section is large but the jets have small
invariant masses.
\end{description}

\nn The analysis proceeded by selecting among all ``reconstructed'' final state
particles the electons and/or positrons with a momentum greater than 30 GeV.
All particles within an angle 10$^0$ along the beam direction were rejected.
Since for the signal, we have incorporated the correlations between all the
particles involved in the process, which have been worked out in detail in (I),
we have full control on the kinematics of the process. This alowed us to
select efficiently the cuts which suppress considerably the background
processes without affecting significantly the signals. \s

\nn For the charged lepton, the following cuts were found to be very efficient
against the background without affecting drastically the signals:
\begin{description}
\item[(1)]  \ Require one and only one $\ee$ pair; this cut makes that one has
not to consider the huge background from single $W$ production, and eliminates
a
very large part of the $\ee Z$ background in which one of the electrons goes
along the beam pipe and therefore escapes detection.
\item[(2)]  \ The invariant mass of the two jets should reconstruct to the $Z$
boson mass, $85 < M_{jj} < 105 $ GeV; the lower bound was set intentionally
high so as to avoid a possibility of confusing a $Z$ with a $W$ boson. This cut
suppresses the $ZZ$ and $\ee Z$ backgrounds by approximately a factor of two.
\item[(3)] \ In the case of $E^-$ production, the momentum of the positron
should
be $|p_{l^+}| > (E_{\rm beam}-M_E^2/4E_{\rm beam})-40$ GeV and the momentum
of the electron $|p_{l^-}| > \frac{2}{3} M_E -133$ GeV, with $M_E$ the
reconstructed mass of the heavy lepton; these kinematic constraints ensure
energy-momentum conservation, with some tolerance for detector resolution
effects. For $E^+$ production, one has to interchange the cuts on the $e^+$ and
$e^-$.
\item[(4)] \ The invariant mass of the $\ee$ pair should be different from
$M_Z$: $|M_{ll} - M_Z| > 12$ GeV; this cut is effective against the $ZZ$
background since the latter process is suppressed by a factor of two.
\item[(5)] \ Cuts on the angle between the $Z$ boson and the initial electron
$\cos\theta_Z > -(M_E+440)/720$ and $\cos\theta_Z < (2100-M_E)/2000$, which are
necessary to reduce the background from $ e^+ e^- \to e^+ e^- Z$; this cut
applies for $E^-$ production, similar ones apply for $E^+$.
\item[(6)] An angular cut $\cos\theta_{ll} < 0.5$ since in the signal, the
two leptons are mostly back--to--back.
\end{description}

\nn In the case of neutral leptons, similar cuts where applied:
\begin{description}
\item[(1)] \ One and only one electron or positron; this cut is very effective
against the background from single $W$ production where the electron goes
along the beam pipe and therefore is undetected.
\item[(2)] \ The invariant mass of the two jets should reconstruct to the $W$
boson mass within 10 GeV, $ 70 < M_{jj} < 90 $ GeV.
\item[(3)] \ The missing momentum is constrained to be $|p_{\nu} | > (E_{\rm
beam} -M_N^2 /4E_{\rm beam})-40$ and the momentum of the charged lepton to be
$|p_{l}| > \frac{2}{3} M_N - 133$.
\item[(4)] \ The invariant mass of the two leptons  should be large: $ M_{ll} >
120$ GeV; this cut is effective against the $WW$ background since in this case
the invariant mass peaks at $M_W$.
\item[(5)] \ The angular cut $\cos\theta_W < 2.58-M_N/240$ reduces the
background from single $W$ production $ e^+ e^- \to e^- \bar{\nu} W$.
\item[(6)] \ An angular cut $\cos\theta_{l\nu} < 0.5$; here also the
two leptons from the signal are mostly back--to--back.
\end{description}

\nn The effects of the previous selection criteria on the main background
processes and on the signals for single production of $N$ and $E$ with masses
of 350 GeV and mixing angles of 0.025 and 0.05 respectively are shown in Table
2 and 3. After these cuts, no events from the $t \bar{t}$ background survived.
In the case of the $\gamma \gamma$ background, because it was not possible to
generate a sufficient number of events, only an estimate of $< 10$ surviving
background events/5 GeV can be given for the mass region $M<250$ GeV. The
background events from vector boson pair production have been suppressed to a
very low level; the background events from single $W$ and $Z$ production are
relatively higher. Note that in the tables, the event number for the signal and
the latter background are of the same order: this is simply due to the fact
that we have taken a large bin [50 GeV] for the reconstructed lepton--jets
invariant mass. Note also that although we have tried to optimize all the above
cuts, we believe that further improvements are still possible. \s

\nn Fig.~9 shows the reconstructed invariant mass histograms for heavy leptons
with masses of 250, 350 and 450 GeV and with mixing angles $\theta_{\rm mix}
=0.05$ in the case of the charged lepton and $\theta_{\rm mix} = 0.025$ for
the heavy neutrino. For these values, one can see that the signal peaks stand
out clearly from the background events, especially for heavy lepton masses not
too close to the total c.m. energy of the collider. Only the background events
from single gauge boson production are relatively important. \s

\nn For smaller mixing angles, the signal cross sections have to be scaled down
correspondingly. For lepton masses of the order of 450 GeV, only slightly
smaller $\theta_{\rm mix}$ values can be probed, while for masses around 350
GeV one can go down by at least a factor of two. The situation is much more
favorable for heavy neutrinos than for charged leptons, the cross section
being one order of magnitude larger. For instance, assuming a mass of 350 GeV
and requiring that the ratio of the signal events to the square root of the
background events be larger than unity, one can probe mixing angles down to
$\theta_{\rm mix} \sim 0.005$ for neutral leptons and $\theta_{\rm mix} \sim
0.03$ for charged leptons. \s

\subsection*{5.~Summary}

In this paper we have studied the production, at future high--energy $\ee$
colliders, of new heavy fermions predicted by several extensions of the
Standard Model. Exploiting the various studies done for a 500 GeV $\ee$
collider, we have used a model detector to analyze the discovery potential of
an $\ee$ linear collider operating at this center of mass energy. \\

\nn If the new fermions have non--zero electromagnetic and weak
charges, they can be pair produced when	their masses are smaller than
the beam energy	of the $\ee$ collider. At a 500~GeV collider, the cross
sections are fairly large being, up to phase space suppression factors, of the
order of the point like QED cross section for muon pair production.
This leads to samples of several thousand events per year for a luminosity of
${\cal L}=10^{33}$ cm$^{-2}$s$^{-1}$. The backgrounds are rather small and
the signals very clean, the large number of events allows to probe
masses	up to practically the kinematical limit of 250 GeV. \\

\nn The	mixing between the new fermions	and the	ordinary fermions of the
Standard Model allows an additional production mechanism: the single production
in association with their light ordinary partners. The cross sections are
suppressed by mixing angle factors and are very small for quarks and for
second and third families of leptons for which the process is mediated only by
$s$--channel $Z$ boson exchange. In the case of the first generation of new
leptons, the reactions are mediated by additional $t$--channel exchanges: $W$
exchange for heavy neutrinos and $Z$ exchange for charged leptons. This
increases the cross sections by several orders of magnitude, and leads to large
numbers of events for not too small mixing angles.
Because	of the purity of the environment of $\ee$ colliders, the signals
are clear and rather easy to separate from backgrounds due to more
conventional processes. Lepton masses close to the total center of mass
energy of the collider can be probed this way.

\newpage

\nn {\large {\bf Table Captions}}

\begin{description}

\item[Tab.~1] Assumed parameters of a standard detector for $\ee$ collisons at
a c.m. energy of 500 GeV.

\item[Tab.~2] Cross sections for the single production of a heavy charged
lepton detected in the process $\ee \ra e^\mp E^\pm \ra \ee jj$ and for the
principal background processes from $\ee \ra ZZ$ and $\ee \ra \ee Z$, after
successive application of selection criteria, as described in the text. For the
heavy charged lepton, a mass of 350 GeV and a mixing angle of 0.05 are assumed.

\item[Tab.~3] Cross sections for the single production of a heavy neutral
lepton detected in the process $\ee \ra \nu_e N \ra \nu_e e jj$ and for the
principal background processes $\ee \ra WW$ and $\ee \ra \nu_e e W$, after
successive application of selection criteria, as described in the text. For the
heavy neutrino, a mass of 350 GeV and a mixing angle of 0.025 are assumed.

\end{description}

\newpage

\nn {\large {\bf Figure Captions}}

\begin{description}

\item[Fig.~1] Feynamn diagrams for the pair production of heavy leptons (a),
for their single production in association with their light partners (b) and
for their decay into a light lepton and a gauge boson which subsequently decays
into two fermions.

\item[Fig.~2] Total cross sections for the pair production of vector and mirror
charged and neutral heavy leptons at a c.m. energy of $\sqrt{s}=500$ GeV, as
a function of the mass.

\item[Fig.~3] Differential cross sections in the pair production of charged
and neutral leptons as a function the scattering angle. The c.m. energy is
$\sqrt{s}=500$ GeV and the mass of the lepton is fixed to 150 GeV.

\item[Fig.~4] Longitudinal and transverse components of the polarization
vectors of the pair produced charged and neutral leptons as a function of the
scattering angle. The lepton masses are fixed to 150 GeV.

\item[Fig.~5] Reconstructed invariant mass histograms for pair produced heavy
charged lepton with masses of 200 GeV in the decays $E^- \ra e^-Z \ra e^-jj$.
For the various cuts and experimental resolutions see the main text.

\item[Fig.~6] Cross sections for the single production of charged and neutral
heavy leptons in association with their light partners at $\sqrt{s}=500$ GeV
as a function of the lepton masses. The left and right--handed mixing angles
are taken to be $\zeta_L , \zeta_R=0.1$.

\item[Fig.~7] Differential cross sections in the single production of charged
and neutral leptons as a function the scattering angle. The c.m. energy is
$\sqrt{s}=500$ GeV, the mass of the lepton is fixed to 400 GeV and the mixing
angles are taken to be $\zeta_L, \zeta_R=0.1$.

\item[Fig.~8] Longitudinal and transverse components of the polarization
vectors of the singly produced charged and neutral heavy leptons as a function
of the scattering angle. The lepton masses masses are fixed to 400 GeV.

\item[Fig.~9] Reconstructed masses of the singly produced heavy leptons in the
processes $\ee \ra e^+E^- \ra e^+e^-Z \ra e^+e^-jj$ (a) and $\ee \ra \nu_e N
\ra \nu_e e^- W \ra \nu_e e^- jj$ (b) for three lepton masses 250, 350 and 450
GeV and for the main backgrounds. The mixing angles have been fixed to
$\theta_{\rm mix}=0.05$ for the charged lepton and $\theta_{\rm mix}=0.025$ for
the neutral lepton. For the various cuts and experimental resolutions see the
main text.

\end{description}

\newpage

\vspace*{3cm}

\begin{table}[hp]
\centering
\begin{tabular}{||c|c|c||} \hline \hline
& & \\
                    & ~~~~~~$\Delta p$; A  ~~~~~~     &   0.001  \\
 Charged            & $\Delta p$; B                   &   0.005  \\
 Particle           & $\cos \theta_{cut}$             &   0.98   \\
 Tracker            & $\Delta \theta$ (mr)            &   2      \\
                    & $\Delta \phi$ (mr)              &   0.5    \\
                    & $ p_T^{min}$ (GeV)              &   0.1    \\

& & \\ \hline
& & \\

                    & $\Delta E$; A                   &   0.08   \\
 Electromagnetic    & $\Delta E$; B                   &   0.02   \\
 Calorimeter        & $\cos \theta_{cut}$             &   0.98   \\
                    & $\Delta \theta$ (mr)            &   2      \\
                    & $\Delta \phi$ (mr)              &   2      \\
& & \\ \hline
& & \\

                    & $\Delta E$; A                   &   0.60   \\
 Hadronic           & $\Delta E$; B                   &   0.02   \\
 Calorimeter
                    & $\Delta \theta$ (mr)            &   10     \\
                    & $\Delta \phi$ (mr)              &   10     \\
& & \\                                                          \hline
\multicolumn{3}{||c||} { }\\
\multicolumn{3}{||c||} { \hspace*{1cm} Momentum resolution $\Delta p/p =
\sqrt{(Ap)^2 + B^2} $ ($p$ in GeV) \hspace*{1cm} }\\
\multicolumn{3}{||c||} { Energy resolution $\Delta E/E = \sqrt{A^2/E + B^2} $
($E$ in GeV) }   \\
\multicolumn{3}{||c||} { }\\
                                                         \hline \hline
\end{tabular}
\caption{}
\label{tab:detector}
\end{table}

\newpage

\begin{table}[hp]
\centering
\begin{tabular}{|c||c|c|c|} \hline
& & & \\
Process & $E^+ e^- + E^- e^+ $ &
$ \ \ e^+ e^- Z \ \ $     &   $ \ \ \  Z  Z \ \ \ $  \\
& & & \\ \hline
& & & \\
 $ \sigma$ [fb]                      &    9.5                     &
4960             &  615         \\
& & & \\ \hline
& & & \\
 $ \times $ B.R.    &    2.19                    &
3470             &    28.8      \\
& & & \\ \hline
& & & \\
  one $\ee$ pair                        &    1.74                    &
93.0            &    23.0      \\
$ 330 < M_E < 370$ GeV           &    1.56                    &
11.7            &     5.30     \\
$  85 < M_Z < 105$ GeV           &    1.41                    &
5.84            &     2.87     \\
$ |M_{ll} - M_Z| > 12$ GeV             &    1.39                    &
5.18            &     1.02     \\
$ \cos \theta_{ll} < 0.5 $              &    1.33                    &
4.32            &     0.56     \\
$ f( M_E, \cos \theta_{Z})$ &    1.30                    &
1.90            &     0.43     \\
 kinem. cuts                           &    1.30                    &
1.55            &     0.39     \\
& & & \\
\hline
\end{tabular}
\caption{}
\label{tab:Ecut}
\end{table}

\bigskip

\begin{table}[hp]
\centering
\begin{tabular}{|c||c|c|c|} \hline
& & & \\
Process  &$ \ \ \ \bar{N} \nu + \bar\nu N  \ \ \ $ &   $ \ \ \ e \nu W   \ \ \
$     &   $  \ \ \ W  W \ \ \ $  \\
& & & \\ \hline
& & & \\
 $ \sigma$ [fb]                             &    490                     &
8610             &   2600       \\
& & & \\ \hline
& & & \\
 $ \times $ B.R.
                                       &    13.7                    &
5823             &   1140       \\
& & & \\ \hline
& & & \\
  one $e$                              &    13.2                    &
198              &    883       \\
$ 330 < M_N < 370$ GeV           &    12.5                    &
11.9             &    100       \\
$  70 < M_W <  90$ GeV           &    12.3                    &
10.3             &     70.3     \\
$  M_{l\nu} > 120 $ GeV                &    11.8                    &
10.0             &     7.93     \\
$ \cos \theta_{l\nu} < 0.5 $              &    11.7                    &
10.0             &     7.80     \\
$ f( M_N, \cos \theta_{Z})$ &    11.7                    &
10.0             &     7.80     \\
 kinem. cuts                           &    11.7                    &
10.0             &     4.13     \\
& & & \\
\hline
 \end{tabular}
\caption{}
\label{tab:Ncut}
\end{table}

\end{document}